\documentclass[aps,prl,amsmath,amssymb,onecolumn,nofootinbib,showpacs,tightenlines,12pt]{revtex4}

\usepackage{epsfig,graphicx}
\usepackage{bm}
\newcommand{\be}{\begin{equation}}
\newcommand{\ee}{\end{equation}}
\newcommand{\bea}{\begin{eqnarray}}
\newcommand{\eea}{\end{eqnarray}}

\begin{document}
\title{ Some notes about the density of  states for a negative
pressure matter }
\author{Nayereh Majd\footnote{ Email: naymajd@ut.ac.ir}}
\affiliation{ Department of Engineering Science, Faculty of
Engineering, University of Tehran, Tehran, PO Box 11155-4563, Iran }
\author{D.Momeni\footnote{Corresponding author}}
\affiliation{Department of Physics,faculty of Basic Sciences,Tarbiat
Moalem University,Karaj, Iran}

  \pacs{3.70.+k, 11.10.-z , 11.10.Gh, 11.10.Hi }

\begin{abstract}
 The main goal of this paper is deriving Density of states
$g(\epsilon)$  (degeneracy function) per volume for an equation of
state (EOS) $p=-\rho$ (we called it dark energy(DE)). We have
concluded that thermodynamic quantities such as pressure and energy
density are simple functions of temperature, fugacity, curvature and
mass of Bosons. Our work has been expressed the origin of some
claims about the negativity of the entropy for the scalar fields
models of DE.
\end{abstract} \maketitle


\section{I: Introduction}
   In recent years some works have been done on the thermal properties of the negative pressure matters, especially on the dark energy. The main assumption of all these models is that the DE is a thermal ensemble at a certain temperature with a specific amount of
  entropy\cite{1}. It would be usable to take temperature as an intrinsic characteristics of DE, namely a barotropic matter with an EOS as $p=f(\rho)$. This fluid must not violate energy conditions. Also we can take both dark matter (DM) and DE as different types of the same matter field with a phase transition in cosmological distances\cite{2}.
     There is not any reasonable statistical description in negative pressure matters .In some of recent papers, thermal properties of dark energy have been discussed in terms of the assumption that the dark energy substance is a thermal ensemble at a certain temperature with an associated thermo
dynamical entropy[1]. It is usually assumed that this temperature is
an intrinsic property of DE rather than the temperature of the heat
bath fixed by surrounding matter.\\
Checking  the second law of thermodynamics in the context of the
black hole physics is another important problem.Also the generalized
second law (GSL) must be satisfied by any physical description of
the DE even in the new scenarios of the modified gravity in the non
relativistic diffeomorphism broken models such as Ho$\breve{r}$ava
proposal[29].If we denote that $S_{tot}$ is the total entropy of a
system,then GSL guaranties that $\dot{S}_{tot}\geq 0$for all
times.Recently Mubasher Jamil , et.al have shown that[26] in an
interactive model for DE , DM and radiation "\emph{the generalized
second law is always and generally valid, independently of the
specific interaction form of the fluid's equation-of-state
parameters and of the background geometry.}".The history of GSL
backs to the Unruh and Wald classical work[28]and Bekenstein works
on the black holes[31].There exists a recondite relation between the
holographic dark energy model and generalized second law of
thermodynamics as in [27] and also in quintom dominated universe
[30].Holographic scenario for DE is so popular and was investigated
by authors both in the context of the usual FRWL cosmology and also
in the modified ones and in a braneworld picture of
universe[25,32,33,36].There is a delicate relation between Bulk
brane interaction , holographic dark energy [34], Gauss Bonnet dark
energy models[35,38],Stringy inspired tachyon model[37],phantom like
regime [39],Observational constraints on holographic dark
energy[40],  Holographic Chaplygin gas model[41], Holographic
Chaplygin DGP Cosmologies[42]and even Holographic Modified
Gravity[43].

    Some years ago Kulikov  and Pronin \cite{3,15}  constructed a simple
      formulation of a  local quantum
  statistics of Bosonic field in a curved background. Indeed they
  explained a simple expression for a grand  canonical thermodynamic
potential density \cite{4} describing locally all
  thermo chemical properties of gases with a curvature dependent
  parts \cite{5}.Later The
high-temperature expansion of the grand thermodynamic potential of
non conformably invariant spin-0 and spin-1/2 gases in an arbitrary
static spacetime with their  boundary is calculated through the
method
conforming from Kirsten [16].\\
  In this report, we have worked on the statistics of a
  negative matter with EOS like DE. At first  we have obtained the
  degeneracy distribution function for massive Bosons in this grand
  ensemble and it has been  shown that the density of states for
  such an
exotic matter is a constant   function of energy .
   We have summed over all energy states once
   more and obtained a closed form for the grand  canonical thermodynamic potential
   density and  then we have inferred the  expressions for the density
of the entropy
   and other important functions. We have also shown that these
   thermodynamic equations are generally related with the fugacity , the Ricci scalar
  and consequently the temperature and
  the chemical potential. The chemical potential has been associated with
a conserved Boson number. We have stated that from a statistical
point of view
   DE violate energy conditions companionship to the claims stated in
the recent
   papers\cite{6}.To shunning from this undesirable result we must
   take both volume and temperature as two dependent variables as Gong
   and collaborations have shown in [6].

  \section{II: Lagrangian of the model and some remarks about the
background metric}
  We begin from Lagrangian of a massive scalar field coupled to a gravity   which we assume that is non negative. Consider the
action\cite{8}
\begin{eqnarray}
S = \int \,\sqrt{-g}
d^{4}x(\frac{1}{2}\phi^{;\mu}\phi_{;\mu}-\frac{1}{2}(m^{2}+\xi R
)\phi^{2})
  \end{eqnarray}
where  $R$  is the  scalar curvature  and we assume that the metric
of the space time  $g_{\mu\nu}$ is static or  slowly time variable
such that we can take the Ricci scalar, time independent. Spherical
symmetry is not essential in this formalism, because we use the
general form for the metric, and only we impose a restriction on the
time variation of the metric. The only essential assumption is on
the surface tension energy portion in EOS. In the case of bulk
matter we ignore the surface effects and all of the thermodynamics
quantities are expressed per unit volume.
  Considering the equilibrium assumption , a static
metric is essential[14]. In this case the background metric
$g_{\mu\nu}$ generated by the mass distribution is assumed to be a
slowly varying function of the time on the inverse temperature
scale[14]. More precisely, we assume
\begin{eqnarray}
\frac{\partial g_{\mu\nu}}{\partial t} \ll
\frac{g_{\mu\nu}}{\beta}\,. \label{eq2005}
\end{eqnarray}
The fluctuation of the metric is due to the energy momentum tensor
fluctuations.
    If we ignore the energy fluctuations, (equilibrium state)
   , we can assume that the metric is static or slow varying with respected to
   the time.
   We neglect the influence of the matter and
radiation and also assume that their interactions with DE are small
and serve only to provide a heat bath at the temperature $T$. The
boundary is a timelike tube which is periodically identified in the
imaginary time direction with the period $\beta$. The necessary path
integral for the construction of the partition function is taken
over asymptotically vanishing fields which are periodic in the
imaginary time t with the period $\beta$. Thus, the functional
integration has been assumed with the periodicity in the imaginary
time and the asymptotic flatness of the metric fields. As we know
that in any static or slowly time varying spacetimes the Ricci
scalar has no dynamics (time dependency) or it's variations with
respect to the time is negligible,but  only in these models we can
write a suitable explicit expression for Green function. On the
other hand the static spacetime implies that the ensemble of Bosons
remains at a definite constant temperature in an equilibrium or a
quasi equilibrium state of the system .If we are  far away from a
static space times, all equations has only a perturbative solution
and it's applications of an expanding universe is doubtful. There is
only some few published works in this topics and finally leads to a
stochastic statistical field theory[20].
\footnote{Stochastic semi classical gravity in the 1990s is a theory
naturally evolved from semi classical gravity in the 1970s and
1980s. In stochastic semiclassical gravity the main object of
interest are the noise kernel, the vacuum expectation value of the
(operator-valued) stress-energy bi-tensor, and the centerpieces
being the (semiclassical) Einstein-Langevin equation. It also brings
out the open system concepts and the statistical and stochastic
contents of the theory such as dissipation, fluctuations, noise and
decoherence. Hu[20] has been described the applications of the
stochastic gravity to the back reaction problems in the cosmology
and the black-hole physics. Further discussions of the ideas and the
ongoing research topics can be found in [21,22,23]}
  The topology of this manifold is considered to be a    $S^{1}\times
M^{3}$ and also we can take $M^{3}$ manifold without any
boundary[17] .

\section{III: thermodynamic potential}
In this section we will analyze the scalar field model with a
conserved charge.
\\
The Lagrangian of the model is:
\begin{eqnarray}
S_{m} = - (\frac{1}{2}) \int d^{4} x \sqrt{g(x)}\Phi^{\ast}(x)(-
\square_{x} + m^{2} + \xi R)\Phi(x)
\end{eqnarray}
Where $\Phi = (\phi_{1},\phi_{2})$ is a doublet of the real fields.
The action written in the terms of real fields will be:
\begin{eqnarray}
S_{m} = -(\frac{1}{2}) \int d^{4} x \sqrt{g(x)} \phi^{a} (x)
(-\square_{x} + m^{2} + \xi R) \phi_{a}(x)
\end{eqnarray}
The total action of the system $"matter+gravity"$ is:
\begin{eqnarray}
S_{tot} = S_{g} + S_{m}
\end{eqnarray}
Now we can write the effective action at finite temperature as:
\begin{eqnarray}
L_{eff}(\beta) = \tilde{L_{g}} - \omega (\beta,\mu,R)
\end{eqnarray}
where $ \tilde{L_{g}} $ is
\begin{eqnarray}\nonumber
\tilde{L_{g}}=L_{g}-\frac{i}{2}\int_{m^2}^{\infty}d m^2 tr
G_{SD}(x,\acute{x})
\end{eqnarray}
The symbol $ tr(...) $  is determined as:
\begin{eqnarray}
tr(...) = \sum_{n \neq 0} \int \frac{d^{3}k}{(2\pi)^{3}}...
\end{eqnarray}
  and $\omega(\beta,\mu,R) $ is the
density of grand thermodynamic potential.
\\
The result $(6)$ may be obtained with the momentum space
representation for the Green's function of a Boson .
\\
In the momentum space representation, the expression for $ L_ {eff}
$ is splited into two parts:
\begin{eqnarray}
L_{eff} = -(\frac{i}{2}) \int^{\infty}_{m^{2}} dm^{2} tr
G(x,\acute{x}) - \omega(\beta,\mu,R)
\end{eqnarray}
The potential $\omega(\beta,R)$ is:
\begin{eqnarray}
\omega (\beta,R) = - (\frac{1}{2}) tr \int^{\infty}_{m^{2}}
dm^{2}\sum^{2}_{j = 0} \sum^{\infty}_{n = 0} \gamma_{j}(R) (-
\frac{\partial}{\partial m^{2}})^{j} \times \int \frac{d^{3}
k}{(2\pi)^{3}} (\omega^{2}_{n} + \epsilon^{2})^{-1}
\\
\nonumber = (\frac{1}{2}) \sum^{2}_{j=0}
\gamma_{j}(R)(-\frac{\partial}{\partial m^{2}})^{j} tr \ln
(\omega^{2}_{n} + \epsilon^{2})
\end{eqnarray}

where $ \omega_{n} = \frac{2\pi n}{\beta}$ and the geometrical
coefficients
  $\gamma_{j}(R)$  can be represented in terms of scalar  R and coupling
  constant $\xi$ as:
\begin{eqnarray}\nonumber
\gamma_{0}(R)=1\\\nonumber
\gamma_{1}(R)=(\frac{1}{6}-\xi)R\\\nonumber
\gamma_{2}(R)=-\frac{1}{180}R_{\mu\nu}R^{\mu\nu}+\frac{1}{180}R_{\mu\nu\sigma\tau}R^{\mu\nu\sigma\tau}
+\frac{1}{6}(\frac{1}{5}-\xi)R_{;\mu}^{\mu}
\end{eqnarray}

For introducing the chemical potential we will change the Matsubara
frequencies[24] $\omega_{n} \rightarrow \omega_{n} + \mu $ and then
thermodynamic potential will be $\omega(\beta,\mu,R).$
\\
Since both positive and negative frequencies are summed, we will
get:
\begin{eqnarray}
\nonumber tr \ln(\omega^{2}_{n} + \epsilon^{2}) \rightarrow tr
\ln[(\omega_{n} + \mu)^{2} +\epsilon^{2}]
\\
= tr \{\ln [ \omega^{2}_{n} + (\epsilon - \mu)^{2}] + \ln
[\omega^{2}_{n} + (\epsilon +\mu)^{2}]\}
\end{eqnarray}
After doing required mode's summation in $(9)$ we will have:
\begin{eqnarray}
\omega(\beta,\mu,R) = \omega_{-}(\beta,\mu,R) +
\omega_{+}(\beta,\mu,R)
\end{eqnarray}
Where:
\begin{eqnarray}
\omega_{-}(\beta,\mu,R) = (\frac{1}{\beta})
\sum^{2}_{j=0}\gamma_{j}(R) ( - \frac{\partial}{\partial m^{2}})^{j}
\ln ( 1-\exp [-\beta(\epsilon - \mu)])
\end{eqnarray}
And:
\begin{eqnarray}
\omega_{+}(\beta,\mu,R) = (\frac{1}{\beta})
\sum^{2}_{j=0}\gamma_{j}(R) ( - \frac{\partial}{\partial m^{2}})^{j}
\ln ( 1- z \exp [-\beta(\epsilon - \mu)])
\end{eqnarray}
So the density of grand thermodynamic potential is the series:
\begin{eqnarray}
\omega(\beta,\mu,R) = \sum^{2}_{j=0} \gamma_{j}(R) b_{j}(\beta m,z)
\end{eqnarray}
Where:
\begin{eqnarray}
\nonumber b_{0}(\beta m,z) = (\frac{1}{\beta}) \ln (1 - z\exp
(-\beta \epsilon));
\\
b_{j}(\beta m,z) = (- \frac{\partial}{\partial m^{2}})^{j}
b_{0}(\beta m,z)
\end{eqnarray}
and the fugacity is $ z = \exp(\beta \mu).$

\section{IV: statistics and thermodynamics of an ideal (non
interactive) Bose gas} The Bose distribution function as the
derivative of the grand thermodynamic potential is given by:
\begin{eqnarray}
n_{\vec{k}} = - \frac{\partial
\omega_{\vec{k}}(\beta,\mu,R)}{\partial \mu}
\end{eqnarray}
For occupation numbers with momentum $ \vec{k} $ we can obtain:
\begin{eqnarray}
n_{\vec{k}} = \frac{1}{(z^{-1} e^{\beta \epsilon_{\vec{k}}}-1
)}B(\beta,R)
\end{eqnarray}
Where the function $ B(\beta,R) $ is described by the formula:
\begin{eqnarray}
B(\beta ,R) = 1 + \gamma_{1}(R)
\frac{\beta}{2\varepsilon_{\vec{k}}}[1 - (1-z\exp^{-\beta
\varepsilon_{\vec{k}}})^{-1}] + ...
\end{eqnarray}
The function $B(\beta,R)$ depends on the curvature, the temperature
and the energy of the Boson[18].
\\
Studying the thermodynamic properties of the Bose gases we will
start with the equation:
\begin{eqnarray}
\omega(\beta,\mu,R) = - (\frac{1}{\beta}) \sum^{2}_{j = 0}
\gamma_{j}(R)(- \frac{\partial}{\partial m^{2}})^{j} \ln (1 - z \exp
[-\beta \epsilon])
\end{eqnarray}
In the non-relativistic limit for the particle energy $\epsilon=
\frac{\vec{k}^{2}}{2m} $ we can derive  from $(19)$ the equation:
\begin{eqnarray}
\omega(\beta,\mu,R) = \sum^{2}_{j=0} \gamma_{j}(R) g_{5/2}(z)(-
\frac{\partial}{\partial m^{2}})^{j} \lambda^{- 3}
\end{eqnarray}
where $\lambda = (2\pi /m T)^{1/2}$ is a wavelength of the particle,
and the function $ g_{5/2}(z) $ has the following form:
\begin{eqnarray}
\nonumber g_{5/2}(z) = \sum^{\infty}_{l=1} \frac{z^{l}}{l^{5/2}}
-\frac{4}{\sqrt{\pi}} \int^{\infty}_{0} x^{2} \ln (1 -
z\exp(-x^{2}))dx
\end{eqnarray}
The average number of the particles in a certain momentum state $
\overrightarrow{k} $ is obtained as the derivative:
\begin{eqnarray}
\nonumber <n_{\overrightarrow{k}}> = - \frac{\partial}{\partial
\mu}\omega(\beta,\mu,R) = \sum^{2}_{j = 0} \gamma_{j}(R)g_{5/2}(z)(-
\frac{\partial}{\partial m^{2}})^{j}(z^{-1} \exp(\beta \epsilon) -1)
\end{eqnarray}
The density of the particles is:
\begin{eqnarray}
n = \lambda^{-3}[1 - \gamma_{1}(R)(3/4m^{2}) -
\gamma_{2}(R)(3/16m^{4})] g_{3/2}(z) + n_{0}
\end{eqnarray}
Where the new function $g_{3/2}(z) $is:
\begin{eqnarray}\nonumber
g_{3/2}(z) = z\frac{\partial}{\partial z}g_{5/2}(z)
\end{eqnarray}
  and
\begin{eqnarray}\nonumber
n_{0} =\frac{z}{1-z}
\end{eqnarray}

    is the average number of the particles with zero momentum. The
functions $ g_{3/2}(z)$ and $g_{5/2}(z)$ are special cases of a more
general class of functions:
\begin{eqnarray}
g_{n}(z) = \sum^{\infty}_{k=1} z^{k}/k^{n}
\end{eqnarray}
In a more simple form the equation $(21)$ may be written as:
\begin{eqnarray}
(n - n_{0})\lambda^{3} = g_{3/2}(z,R)
\end{eqnarray}
Where:
\begin{eqnarray}
g_{3/2}(z,R) = [1 - \alpha \frac{R}{m^{2}} + ...] g_{3/2}(z)
\end{eqnarray}
is a function which depends on the curvature, and $\alpha$ is a
numerical parameter.
\\
The equation $(23)$ connects four values: fugacity, temperature,
density of the particles and curvature.
\section {V: calculating the density of states for EOS $p=-\rho$}
   The density of grand thermodynamic potential is :

\begin{eqnarray}
\omega(\beta,\mu,R)=\frac{\Omega}{V}-\frac{1}{\beta}\sum^{2}_{j=0}\gamma_{j}(R)(-\frac{\partial}{\partial
m^{2}})^{j}\sum_{\epsilon}\ln(1-z exp(-\beta\epsilon)),
\end{eqnarray}
  where $\beta=1/T$.

   We can write the sum over the energy states as an integration over
energy values by considering it's occupation number function
(Degeneracy of the energy levels)  $g(\epsilon)$  where:
\begin{eqnarray}\nonumber
g( \epsilon)d\epsilon=\frac{V}{(2\pi)^{3}}d^{3}k
\end{eqnarray}
  We know that for photons, phonons and non-relativistic particles
with spin $s$ this function is(in Geometrical units where:
$\hbar=c=1$)  \cite{11} :
\begin{eqnarray}
\nonumber&& g( \epsilon)=\frac{V\epsilon^{2}}{(2\pi)}(photons)
\\\nonumber&& g(
\epsilon)=\frac{9N\epsilon^{2}}{\omega_{D}^{3}}(phonons)
\\\nonumber&&
g(\epsilon)=\frac{(2s+1)Vm^{3/2}\epsilon^{1/2}}{\sqrt{2}\pi^{2}}(Non
relativistic Bose gas)
\end{eqnarray}

  We now review some relations between the Grand Thermodynamic
   Potential of the system  $\Omega(\beta,z,V)$,   the Helmholtz Free energy
   (F), the entropy (S)and the pressure (P)  \cite{11}:
    \begin{eqnarray}\label{aa}
\nonumber&&F=U-TS\\
\nonumber&&U=\Omega+\mu N
+TS\\ \nonumber&&PV=-\Omega\\
\nonumber&&S=-\frac{\partial\Omega}{\partial
T}\\
\nonumber&&\Omega=\beta(Ln(Z(\mu,T,N))\\
\nonumber&&U=-(\frac{\partial Ln(Z(\mu,T,N))}{\partial\beta}),
\end{eqnarray}
  where in these relations  z is called fugacity :
\begin{eqnarray}\label{aa}
\nonumber&&z=e^{\beta\mu}
\end{eqnarray}
  Since now we named the density of the Grand potential of the  system
as   $ \omega$   then we can write next equations for the density (
  per volume) of them as:
\begin{eqnarray}\label{aa}
f=\rho-T s\\\rho=\omega+\mu n +T s\\P=-\omega
\end{eqnarray}
  The main goal of this paper is deriving degeneracy function per
volume for an EOS $p=-\rho$, and then if it is the well behavior
function, we can fit it with an interaction inside a Boson gas, then
the generation function of both systems will become the same. Our
result for the dispersion relation has been shown that there is the
$k^3$ dependence of the energy. This dependence can be the main
effective term of the interaction in a definite  density and
temperature, and also like the logarithmic dependence of the
dispersion relation strength quark matter (SQM); the $k^3$ term can
be dominant in the some regions[19].


   Remembering previous relations together with EOS   $p=-\rho$
    we lead to the next first order partial differential
  equation for $\omega$.
\begin{eqnarray}
\mu\frac{\partial\omega}{\partial\mu}=\beta\frac{\partial\omega}{\partial\beta}
\end{eqnarray}
now we note here that the general solution for above equation in
which
\begin{eqnarray}
\omega(\mu,\beta)=\sum _{a}c_{a}(\mu\beta)^{a}
\end{eqnarray}
we have a simple boundary condition which is concluded from
thermodynamics: the entropy of a system must be vanish at absolute
zero temperature  $\beta$. that is we must have:
\begin{eqnarray}\nonumber
\lim_{\beta\rightarrow \infty}
(\beta^{2}\frac{\partial\omega}{\partial\beta})=0
\end{eqnarray}
  Thus we must limit ourselves only to  $a<-1$. But the coefficients
remain undetermined. Thus we take another endeavor to solve this
equation. We substitute ( 25) in (29) and take derivatives and
equalize from both sides all terms which have the same order of
$m^{2}$ (i.e.; the coefficients of $(-\frac{\partial}{\partial
m^{2}})^{j}, j=0,1,2$
  We have:
\begin{eqnarray}
\int^{\infty}_{0}\frac{\epsilon
g(\epsilon)}{e^{-\beta(\mu-\epsilon)}-1}d\epsilon-\frac{1}{\beta}\int^{\infty}_{0}\ln(1-ze^{-\beta\epsilon})g(\epsilon)d\epsilon
\end{eqnarray}
    We assume that $ g(\epsilon)=\frac{d\sigma(\epsilon)}{d\epsilon}$
  and also we suppose that   $\sigma(\epsilon)$ is a well behavior
  function of  energy   $\epsilon$  at the ground state energy level
  $\epsilon=0$. After integrating part by part and expanding both
  sides in terms of  z  ($0<z<1$) and set equal all terms of same z
  power we have:
\begin{eqnarray}
\int^{\infty}_{0}e^{-\beta
n\epsilon}[\sigma(\epsilon)-\epsilon\frac{d\sigma}{d\epsilon}]d\epsilon=\frac{\sigma(0)}{\beta
n}, n\geq 1
\end{eqnarray}
   We can define Laplace transformation of the function $\sigma(\epsilon)$ as a
  function of parameter  $s=\beta n$ by:
\begin{eqnarray}
\Sigma(s)=\int^{\infty}_{0}e^{-s \epsilon}\sigma(\epsilon)d\epsilon
\end{eqnarray}
  We arrive at the next differential equation for the transformed function:
\begin{eqnarray}
\Sigma(s)+\frac{\partial} {\partial
s}(s\Sigma(s)-\sigma(0))=\frac{\sigma(0)}{s}
\end{eqnarray}
  Which has the solution:
\begin{eqnarray}
\Sigma(s)=\frac{\sigma(0)}{s}+\frac{C_{1}}{s^2}
\end{eqnarray}

   In this function   $C_ {1} $ is an unknown constant that can be
determined using initial conditions. But we take it as a constant
and find it's value some later. Because we solve the partial
differential equation (29), then the result can be a function of
other quantities like the mass of particles. The mass is not an
independent thermodynamic variable. If we assume the mass as a
thermodynamic variable, then the derivative of mass with respected
to the density, the temperature or the chemical potential must be
occurred in the fundamental thermodynamic relations [19].

  By taking inverse Laplace transformation from (35)we obtain:
\begin{eqnarray}
\sigma(\epsilon)=\sigma(0)+C_{1}\epsilon
\end{eqnarray}
  We mention here that for a Boson's system the chemical potential is
always negative namely $-\infty<\mu\leq0$. Since
$g(\epsilon)=\frac{d\sigma(\epsilon)}{d\epsilon}$ we finally obtain:
\begin{eqnarray}
g(\epsilon)=C_{1}
\end{eqnarray}
  Substituting (37) in definition of
   $g(\epsilon)$ \footnote{Note that V is disappeared since we
calculate density.}
   we lead to the
  following dispersion relation for DE:
\begin{eqnarray}
k^{3}=6\pi^2 C_{1}\epsilon-6\pi^2 C_{2}
\end{eqnarray}
We know that this is not a dispersion relation for a non interactive
matter field.

  But there is no simple system which has such
distribution. If we assume that the energy levels of such Bosonic
system must be positive and the momentum of any particle also be
positive, then the energy levels of this system must satisfy  the
following inequality:
\begin{eqnarray}
\epsilon\geq \frac{C_{2}}{C_{1}}
\end{eqnarray}
Without any dismay from loss of generality we can take $C_ {2} =0$
.With this choice we accept that our ground state energy level is
located at $\epsilon=0$. Another pickup is tagging this energy level
a non zero value. This situation could be solved by a shift of any
energy value up to this value.

  Substituting (37) in (25) and making use from:
\begin{eqnarray}\nonumber
Poly\log(2,z)=\sum^{\infty}_{n=1}\frac{z^n}{n^2},
\end{eqnarray}
  finally we obtain next function for the density of  grand thermodynamic
  potential:
\begin{eqnarray}
\omega(\mu,R,\beta)=\frac{1}{\beta^2}Polylog(2,z)\sum^{2}_{j=0}a_{j}(R)(-\frac{\partial}{\partial
m^2})^jC_{1}
\end{eqnarray}
Now we calculate the  density of the average of the energy .We can
label it as $\rho_{\Lambda}$ which is:
\begin{eqnarray}\nonumber
\rho_{\Lambda}=\frac{\bar{E}}{V}=\int_{0}^{\infty}\frac{g(\epsilon)\epsilon}{e^{-\beta(\mu-\epsilon)}-1}d\epsilon
\end{eqnarray}
After substituting (37) in it and doing the integration we obtain:
\begin{eqnarray}
\rho_{\Lambda}=\frac{C_{1}}{\beta^2}Polylog(2,z)
\end{eqnarray}
The  mean value of the number of the particles in the unit volume of
the system is:
\begin{eqnarray}
n=\frac{\bar{N}}{V}=\int_{0}^{\infty}\frac{g(\epsilon)}{e^{-\beta(\mu-\epsilon)}-1}d\epsilon=-\frac{C_{1}}{\beta}\ln(
1-z)
\end{eqnarray}
Since $0<z<1$ the value of $n$ is not negative.  Using
$s=\frac{S}{V}=-\frac{\partial\omega}{\partial T}$ we have:
\begin{eqnarray}
s=-\frac{2}{\beta}Poly\log(2,z)\sum^{2}_{j=0}a_{j}(R)(-\frac{\partial}{\partial
m^2})^jC_{1}
\end{eqnarray}
Because the mass dependence of $g(\epsilon)$ is undetermined , so we
can not conclude strongly  about the sign of entropy. It can be
negative or positive [14].

Negative entropy is problematic if we accept that the entropy is in
 the association with the measure of the number of microstates in
statistical mechanics. The intuition of the statistical mechanics
requires that the entropy of all physical components have to  be
positive. Besides if we consider the universe as a thermodynamic
system, the total entropy of the universe including DE and DM should
satisfy the second law of thermodynamics. The GSL   for phantom and
non-phantom DE has been explored in [12]. It was found that the GSL
can be protected in the universe with DE. The GSL of the universe
with DE has been investigated in [13,25,26,27] as well. \footnote{We
may think the DE temperature is equal or proportional to the horizon
temperature $T_{H}$.} ,and also the equation of state of the DE is
uniquely determined and the phantom entropy is negative [13]. For
the phantom,when we running to the  negative entropy problem, the
GSL is violated.\footnote{ It is more realistic to consider that the
physical volume and the temperature of the universe are related,
since in the general situation they both depend on the scale factor
a(t).It was found that the apparent horizon is a good boundary for
keeping thermodynamics' laws [13].} Considering the apparent horizon
as the physical boundary of the universe, it was found that both the
temperature and the entropy can be positive for the DE, including
phantom.In the general framework that we discussed, where the
cosmological metric has a slow varying time dependency, we can
assume that the time contiguity of the  volume and the temperature
is venial and for this reason we have obtained a negative entropy.
We must note here that if in an ensemble of particles the volume and
temperature were non static, the statistical behavior of this system
is not in the context of the equilibrium thermodynamics.\footnote{
As in a FRW standard model of cosmology, since our background metric
is not static and thereupon the apparent horizon is time variable,
the construction of a local quantum statistical  and counting all
microstates  must affect the negative entropy and we can deduce a
statistical proof for non negativity and non violation of GSL like
Gong and collaborations proof [6].}

\section{VI: conclusions}
We have derived a grand canonical partition function for a
  DE's like EOS. The thermodynamic equations are generally
expressed in terms  of three  variables: the temperature $T$ ,
chemical potential $\mu$ associated with a conserved particle number
N, and scalar curvature $R$.  We also reviewed some known results
for an ideal Bose gas in curved spacetime. Comparing our results
with an ideal gas, we have concluded
  that there is a significance generic difference between the
statistical distribution of negative pressure matters and ideal
gases. Specifically , the dispersion relation for a dark energy like
EOS is non quadratic (in contradiction with a non interactive ideal
Bose gas)and comparing this dispersion function with the statistical
model of SQM  we observe that also in SQM systems this dispersion
relation is valid in some physically accessible region of the
system[19].

Also we have derived the mean value of the energy of the system in
terms of the curvature . We have shown that if a matter field
(Boson) obeys  an EOS like DE, and in cosmological language violates
the well known energy conditions in General relativity , the
degeneracy function of this gas must be an implicit constant
function of the energy. This function implicitly is dependent on the
mass of the particles which is not a thermodynamic quantity and
remains as an undetermined function in this formula.

The derived thermodynamic equations has been shown that the entropy
may be positive or negative. As it was stated in [6], this
negativity of the entropy is hidden behind this significance
assumption that the volume and the temperature are independent. But
we know that in an expanding FRW universe both of them are functions
of time. Hence if we assume that our metric is quasi static or has a
negligible dynamics, any phantom field of DE with an intelligible
statistical testimony has a negative entropy. If we want to
generalize our argument to a realistic cosmological case, we must
work in a non equilibrium statistical regime.
\section{ Acknowledgment}
The authors thank S. Nojiri , S. D .Odintsov and Neven Bilic for
useful comments and valuable suggestions. Finally, the editor of
IJMPE, and the anonymous referees made excellent observations and
suggestions which resulted in substantial improvements of the
presentation and the results.Nayereh Majd would like to thank
University of Tehran for supporting this project under the grants
provided by the research council, and especially Department of
Engineering Science for its financially supports.

\end{document}